\documentclass[a4paper]{jpconf}
\usepackage{color}
\usepackage{graphicx}
\usepackage{latexsym}
\begin{document}
\title{Improved source localization with LIGO India}

\author{Stephen Fairhurst}

\address{Cardiff University}

\ead{stephen.fairhurst@astro.cf.ac.uk}

\begin{abstract}
 
A global network of advanced gravitational wave interferometric detectors is under 
construction.  These detectors will offer an order of magnitude improvement in 
sensitivity over the initial detectors and will usher in the era of gravitational wave 
astronomy.  Gravitational wave sources can be localized on the sky, primarily 
through time of arrival measurements at the detectors, allowing for followup 
observations by telescopes.  In this paper, we evaluate the improvement in sky 
localization obtained by relocating one of the Advanced LIGO detectors to India.
\end{abstract}

There is currently a proposal to situate one of the Advanced LIGO \cite{Harry:2010} 
detectors in India \cite{Iyer:2011wb}.  This would entail installing one detector at 
each at Livingston, LA and Hanford, WA(rather than two at Hanford) and the third 
detector in India.  The global network will be augmented by the Advanced Virgo 
detector in Cascina, Italy \cite{adV} and the KAGRA detector in Kamioka, Japan \cite
{kagra}.  One of the main motivations for the move to India is the improvement in 
localization of sources afforded by a network with an additional site remote from the 
other sites.  Localization of gravitational wave sources to (relatively) small regions of 
the sky will enable wide field electromagnetic telescopes, such as LOFAR \cite{lofar}, 
Palomar Transient Factory \cite{Rau:2009yx}, Pan Starrs \cite{panstarrs} and 
SkyMapper \cite{skymapper} to perform followup observations of gravitational wave 
events.  The joint electromagnetic and gravitational wave observation of sources 
willprovide new insight into the central engines powering electromagnetic transients 
\cite{nakar07, Ott} and allow for independent measurements of distance and redshift 
sources \cite{Schutz:1986gp}.

In this paper, we investigate in detail the improvement in localization
afforded by the installation of a LIGO India detector.  At this stage, a
site for a detector in India has not been fixed so we consider two
locations.  The first, as introduced by Schutz \cite{Schutz:2011fn},
has the detector placed at the location of the Giant Metre-wave Radio
Telescope (GMRT) \cite{GMRT} at 74 deg  02' 59" E 19 deg 05' 47" N.  The
second is a site close to Bangalore, at 76 deg 26' E 14 deg 14' N, in an
area that is seismically quiet \cite{Iyer:2011}.  In both cases, the
arms are chosen (arbitrarily) to be oriented North and West, although 
the orientation of the detector will not greatly affect the localization 
\cite{LIGOSouth, Veitch:2012df} .  
Examining two locations gives some insight into the effect of moving
the detector.  Of course a more detailed study (such as the one
described in \cite{LIGOSouth}) would be required before fixing the
location and orientation of a detector in India.

The primary method by which a gravitational wave source will be
localized on the sky is by triangulation using measured time
delays between the sites \cite{Fairhurst:2009tc}.  Thus the localization
accuracy will be dominated by the timing accuracy in the detectors as
well as the separation between sites.  In \cite{Fairhurst:2009tc}, the
timing accuracy in a given detector was shown to be 
\begin{equation}\label{eq:sigma_t}
  \sigma_{t} = \frac{1}{2 \pi \rho \sigma_{f}} \, .
\end{equation}
Timing accuracy is inversely proportional to both the SNR $\rho$ and
effective bandwidth $\sigma_f$ of the source, defined as 
\begin{eqnarray}
  \rho^{2} &=& 4 \int_{0}^{\infty} \frac{| h(f) |^{2}}{S(f)} df \, , 
    \nonumber \\
  \sigma_{f}^{2} &=& 
    \left(\frac{4}{\rho^{2}} \int_{0}^{\infty} f^2 
    \frac{|h(f)|^2}{S(f)} df \right) - 
    \left(\frac{4}{\rho^{2}} \int_{0}^{\infty} f 
    \frac{|h(f)|^2}{S(f)} df \right)^{2} \, ,
\end{eqnarray}
where $h(f)$ is the Fourier transform of the signal and $S(f)$ denotes
the one sided noise power spectral density of the detector.  The
approximations used to obtain equation (\ref{eq:sigma_t}) break down at
low SNR, where second order effects become important
\cite{Vitale:2010fg}.  Also, we have neglected the effect of correlation
between timing and other parameters.  These correlations will affect the
observed end time in all sites in a similar way and will thus not have a
significant effect on the recorded time delays between sites.  

We consider localization of a source using only observed time delays between the 
sites.  From a set of observed time delays, the sky location of the source can be 
inferred.  The errors on the arrival times determine the uncertainties in the location.  
For two sites, a single time delay suffices to localize the source to a ring in the sky.  
In principle, it may be possible to break the degeneracy on the ring based on 
additional information, such as spin or other features in the signal
\cite{vanderSluys:2007st}.  Even then the localization area will be large, and we will not consider 
localization regions for a two detector network here.  For three sites, the observed 
time delays lead to two possible locations of the source which are mirror images in 
the plane of the detectors.  In many cases, the observed amplitudes of the signal 
in the three detectors will enable the correct choice of a single location, although this 
is not always possible \cite{Virgo:2011aa, Veitch:2012df}. Here, we assume that the 
degeneracy 
can be broken and for each source consider only the 90\% confidence localization 
region around the true sky location.

There are numerous proposed configurations for the advanced detectors
\cite{Harry:2010, aligo}, and it is likely that several configurations
will be used as the sensitivity of the detectors develops.  For this
work, we are primarily interested in the sensitivity of the detectors
and the frequency bandwidth of the signal.  For simplicity, we will
restrict attention to Binary Neutron Star (BNS) systems and characterise
the sensitivity by the \textit{BNS horizon} --- the distance at which an
optimally oriented and located BNS system would be observed with a
signal to noise ratio of 8.  The \textit{BNS range} ---the volume and
orientation averaged distance at which a BNS gives SNR 8 --- is a factor
of 2.26 smaller than the \textit{horizon}.  In \cite{Fairhurst:2010is},
we listed the sensitivity and frequency bandwidth of a number of
different configurations of the advanced detectors with various
sensitivities and bandwidths.  In this paper, we use a fiducial detector
with a BNS range of 160 Mpc (360 Mpc horizon) and a frequency bandwidth
of 100 Hz.   It is important to note that the results obtained are just
illustrative of what might be achievable by an advanced detector
network.  The actual localization of sources will depend upon the
sensitivities of the detectors and may also be influenced by
non-stationary features in the noise.  Furthermore, it is very likely
that the source localization of the detector networks will evolve over
time as the detectors approach their final, design sensitivities. 

In \cite{Fairhurst:2010is} we introduced several characterizations of detector
networks with regard to localization.  Here we present the results for
networks containing a detector in India.  Where appropriate, we also show
the comparable network with two detectors in Hanford, instead of one
Hanford and one India, to give a sense of the improvement.  
First, we examine the localization for a source at a fixed distance.  We
simulate the signal for a BNS merger, oriented face on, at a distance of
160 Mpc (equal to the BNS range of a single detector) and compute the
90\% localization area.  We require the expected network SNR of the
source to be greater than 12 and require a single detector SNR above 5
in at least two detectors.   This excludes regions of the
sky where a source at 160 Mpc would not be confidently detected.  These
thresholds are motivated by results of recent searches of LIGO and Virgo
data \cite{2012PhRvD..85h2002A}.

\begin{figure}[t]
\includegraphics[width=.49\textwidth]{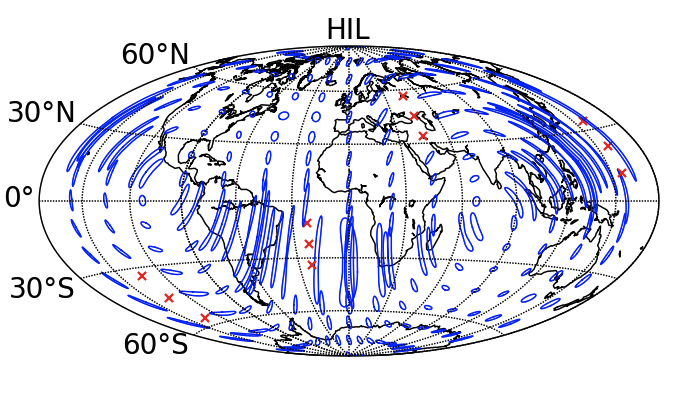} 
\includegraphics[width=.49\textwidth]{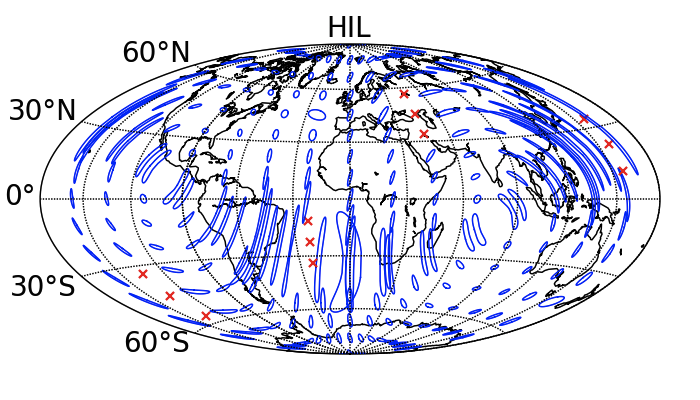}
\caption{Localization with a three site LIGO network.  The plots
show the 90\% localization ellipses for a face on BNS merger at 160
Mpc. A $\times$ is plotted at points where the network would not
confidently detect the system. The left plot shows the localization
ellipses with a detector in India at the location of GMRT.  The right
plot shows localization for a site close to Bangalore.  The qualitative
features of the plots are similar but the localization ellipses do vary
substantially in some parts of the sky.  This is most evident close to
the plane defined by the three detectors.  Moving the detector
changes this plane and consequently the regions of the sky for which
sources cannot be well localized.}
\label{fig:ligo_loc}
\end{figure}

In Figure \ref{fig:ligo_loc} we show the localization of face on BNS sources 
at 160 Mpc that is
possible using three LIGO detectors located at Hanford, Livingston and
India.  We show the results for the two different LIGO India locations
described above.  The detector orientation will not have any
effect on these results as we are considering only face on systems which
give a circularly polarized gravitational wave signal, although the orientation will
have an effect on localization of generic binary systems.  In both cases,
there are large areas of the sky for which the localization is good, but
there is a clear band where localization is rather poor and the ellipses
become elongated.  This corresponds to sources lying in (or close to)
the plane formed by the three detectors.  For sources close to the
plane, the localization out of the plane is poor as a large change in
position corresponds to only a small change in time delays between sites.
We have not shown the corresponding plots for a
Hanford--Hanford--Livingston network as the two site network would only
serve to localize sources to a ring on the sky.  

\begin{figure}[t]
\centering
\includegraphics[width=.49\textwidth]{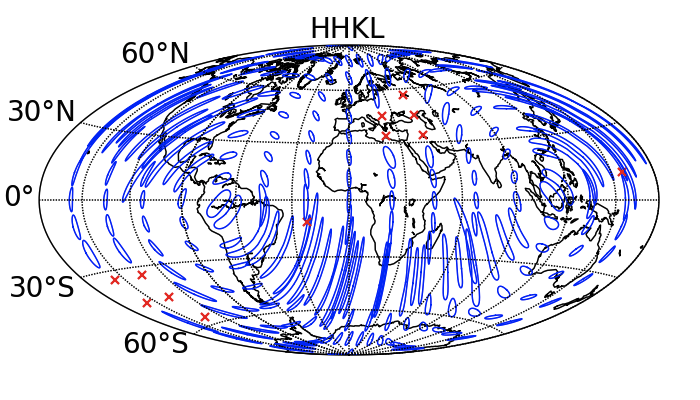}
\includegraphics[width=.49\textwidth]{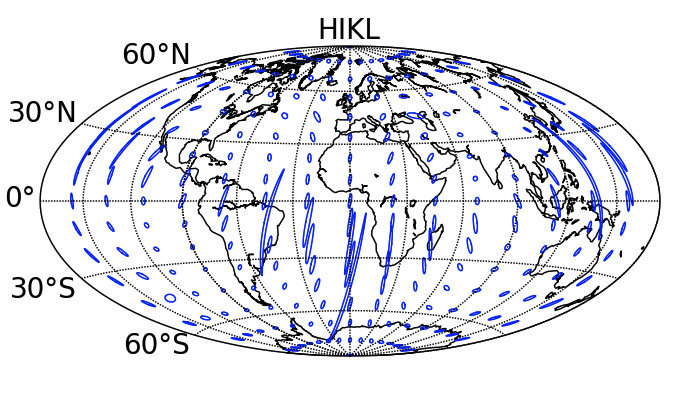}
\includegraphics[width=.49\textwidth]{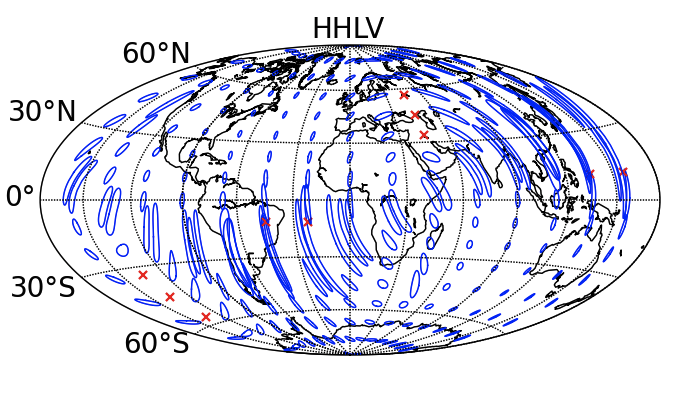}
\includegraphics[width=.49\textwidth]{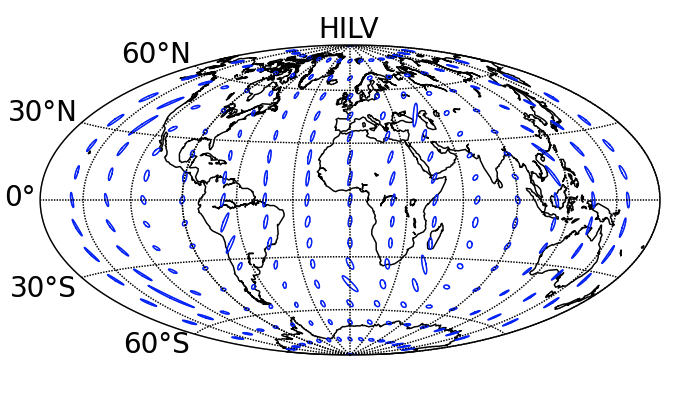}
\includegraphics[width=.49\textwidth]{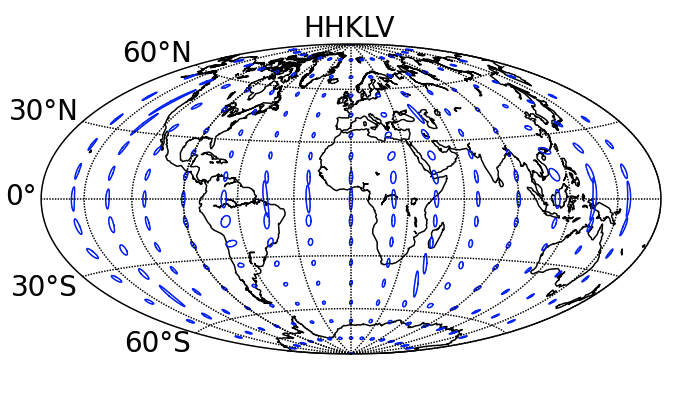}
\includegraphics[width=.49\textwidth]{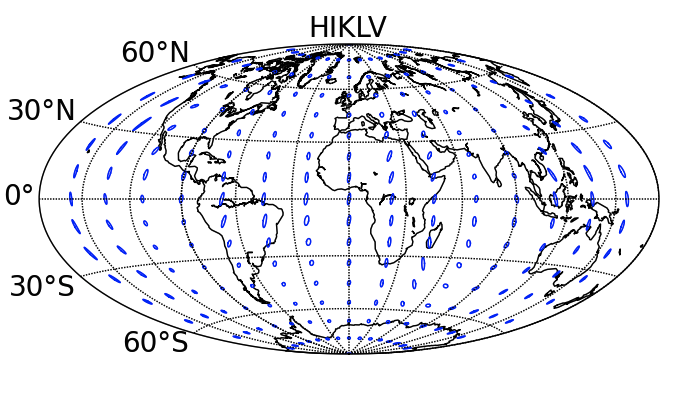}
\caption{The localization accuracy for face on BNS at 160 Mpc in various
networks of advanced detecors.  The ellipses contain the 90\%
localization regions for sources from various points in the sky.  A
$\times$  is plotted at points where the network would not confidently
detect the system.  The plots show the localization for six different
networks: Hanford--Hanford--KAGRA--Livingston (HHKL);
Hanford--India--KAGRA--Livingston (HIKL);
Hanford--Hanford--Livingston--Virgo (HHLV);
Hanford--India--Livingston--Virgo (HILV);
Hanford--Hanford--KAGRA--Livingston--Virgo (HHKLV);
Hanford--India--KAGRA--Livingston--Virgo (HIKLV).}
\label{fig:network_localization}
\end{figure}

Next, we consider the global network comprising three LIGO detectors,
the Virgo detector and the KAGRA detector.  In Figure
\ref{fig:network_localization} we show the localization afforded by a
network comprising four or five detectors (three LIGO plus one or both
of Virgo and KAGRA).  For each network, we show localizations with
either two US LIGO sites (with two Hanford detectors) or three LIGO
sites including a LIGO India detector located near Bangalore.  The three
site networks (HHKL and HHLV) provide comparable localizations to the
three site LIGO network HIL in that there are regions of the sky where
sources are well localized, other regions where the localization is poor
and even a few areas where the source would not be confidently detected.
The results for HHKL show poorer localization than the other three site
networks as these sites are rather close to being co-linear (as
discussed in \cite{Fairhurst:2010is}) so there is only really one good
direction of localization.  The four site networks show good
localization over the majority of the sky with slightly worse
performance for the HIKL network as the four sites are nearly co-planar
and sources lying close to this plane are not well localized.  The five
site network shows good localization over the entire sky.

\begin{figure}[t]
\centering
\includegraphics[width=.49\textwidth]{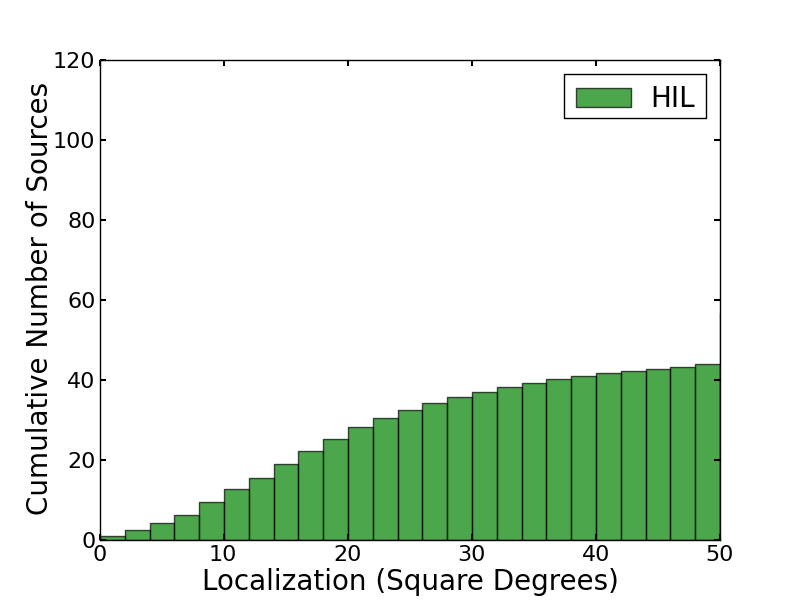}
\includegraphics[width=.49\textwidth]{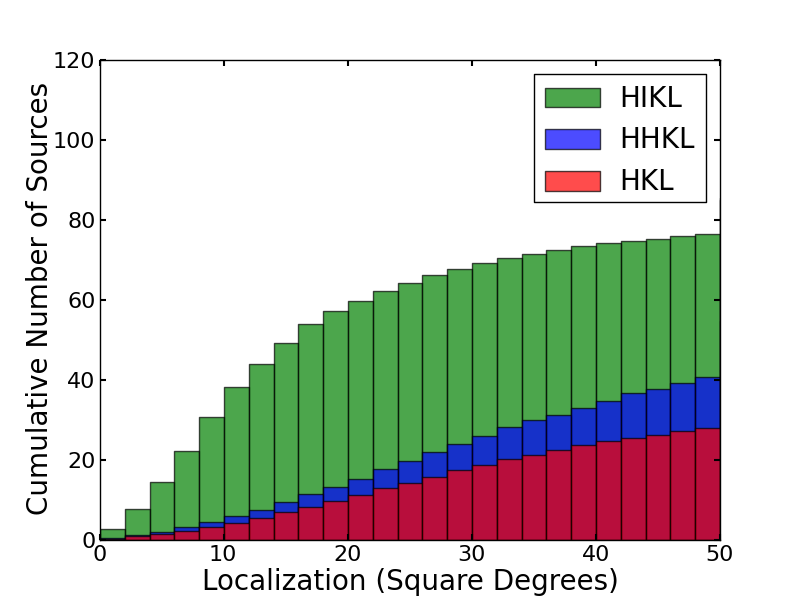}
\includegraphics[width=.49\textwidth]{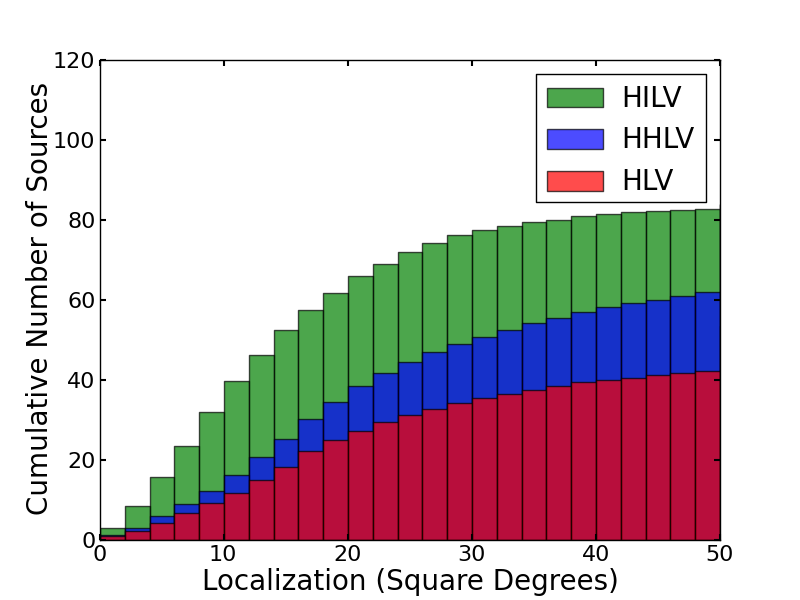}
\includegraphics[width=.49\textwidth]{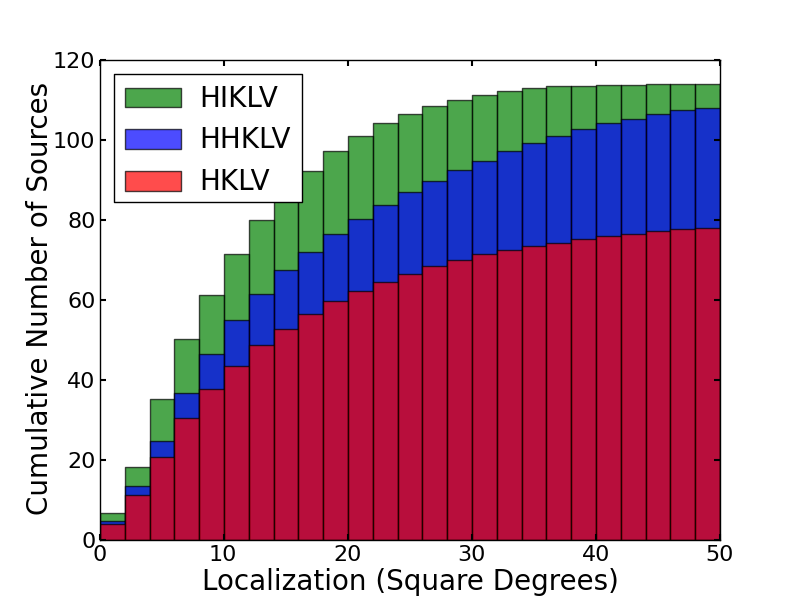}
\caption{The number of sources localized within a given area in the sky
for various networks of detectors.  LIGO only networks (upper left) ---
HIL shown, HHL and HL omitted as no sources are well localized.
LIGO-KAGRA networks (upper right) --- HIKL, HHKL, HKL shown (from top to bottom).  LIGO-Virgo
networks (upper right) --- HILV, HHLV and HLV shown.  LIGO-KAGRA-Virgo
networks (lower left) --- HIKLV, HHLKV and HKLV shown.  In all cases,
the presence of a detector in India (rather than two at Hanford) improves
the source localization dramatically, although it does not significantly
affect the total number of observable signals (not shown on the plots).
A single Hanford detector (and no detector in India) leads to a 25-40\%
reduction in well localized sources, and indeed in the overall number of
observed sources.  All results assume a rate of 40 events per year at
SNR 8 or higher in a single detector.}  
\label{fig:network_hist}
\end{figure}

Next, we would like to investigate the distribution of 90\%
localization areas for a realistic catalogue of detectable sources.  We
choose the sources to be uniformly distributed in volume and source
orientation and impose the same thresholding criteria introduced
above to restrict to detectable signals.  The rate of BNS mergers in the
nearby universe is uncertain to about two orders of magnitude
\cite{Abadie:2010cfa}.  For illustrative purposes, we choose a rate that
would give 40 sources per year with an SNR greater than 8 in a single
detector.  This corresponds to the ``realistic'' astrophysical rate of
\cite{Abadie:2010cfa}.  Since we use the same rate for all the networks,
the results are directly comparable, but the precise numbers should not
be interpreted too strictly as the true rate could reasonably be an order of
magnitude larger or smaller than the rate used here.

The number of sources that will be detected by the various networks
increases significantly with the addition of a detector, but varies
little between the networks.  With our assumed rate, the three detector
networks would observe 55-60 sources annually; four detector networks
82-87; five detector networks 110-115. However, the fraction of sources
that are well localized will vary significantly with detector locations.
Figure \ref{fig:network_hist} shows how well sources are localized with
the various three, four and five detector networks.  In all cases, there
is a clear improvement in localization between a network with a LIGO
detector in India over the corresponding network with two detectors at
Hanford.  This is particularly evident for a LIGO only network, where the
(two-)Hanford--Livingston network is incapable of localizing sources to
better than a ring on the sky while almost half of the sources observed 
by Hanford--Livingston--India are localized
to within $20 \deg^2$.  For the four detector networks, an detector in
India provides two or three times as many sources localized within
$20 \deg^{2}$ compared to having two LIGO detectors at Hanford.  The
five site network opens the possibility of the majority of sources to be
localized within $10 \deg^{2}$.  These localizations are of interest as
the current and  future generations of wide field, transient telescopes
have field of views from $3$ to $20 \deg^2$ \cite{lofar, Rau:2009yx,
panstarrs, skymapper}.  Thus, it will be feasible to perform follow-up
observations of the majority of gravitational wave sources.  

In the long term, we expect to be in a position where five advanced
gravitational wave detectors are operating at their design sensitivity.
Even at this stage, there will be a considerable advantage to having
each detector at a unique site.  It seems unreasonable to assume that
the detectors will operate with 100\% duty cycle --- both due to the
need to perform maintenance on the detectors, and the inevitable
disturbances from local environmental factors.  Indeed, it is typical to
assume that each detector will have around an 80\% duty cycle
\cite{Reitze:Bf7sDr2U}, with the downtime in the two Hanford detectors
being highly correlated.  The five site network provides considerable
redundancy and, assuming 80\% duty cycle in each detector, would give
nearly 75\% duty cycle for a four or more detector network and 95\% duty
cycle for a three or more detector network.  This compares to a 40\%
four site duty cycle and 80\% three or more site duty cycle for the
global network with two LIGO Hanford detectors.

Finally, we must consider the fact that the re-location of a detector to
India will delay its installation by several years \cite{Iyer:2011wb}.
Consequently, there will be a time when the network is operating with
only two Advanced LIGO detectors.  In Figure \ref{fig:network_hist}, we
also show localization of sources using networks with one LIGO detector
at Hanford and one at Livingston, but no LIGO India detector.  These
show a reduction in the number of well localized sources by between 25
and 40\%, and indeed a reduction in the overall number of observed
sources by a similar amount.  For example, with the realistic
astrophysical rate and the above duty cycle assumptions, each year the
HLV network provides 13 sources localized to better than $20 \deg^{2}$
and 40 total sources compared to 18 well localized and 60 total sources
for the HHLV network.  Thus, the relocation of a detector to India would
cause a reduction in the number of sources detected (and localized) in
the short term, but lead to significant long term benefits.  

In summary, we have discussed in detail how the re-location of one of
the LIGO detectors to India provides a significant improvement to the
ability of the global gravitational wave network to localize sources, as
well as providing a greater redundancy to the network.  This improvement
will come at the cost of a delay in the completion of the third Advanced
LIGO detector, but the long term benefits make this worthwhile.  The
ability of four and five site gravitational wave detector networks to
accurately localize sources will be critical to enabling gravitational
wave observations to play their full part in the upcoming era of
multi-messenger astronomy.  

\section*{Acknowledgements}

This work has benefitted from discussions with numerious colleagues
including Duncan Brown, Ray Frey, Bala Iyer, Dave Reitze, Bangalore
Sathyaprakash, Patrick Sutton and Stan Whitcomb.  This research was made
possible thanks to support from the Royal Society.

\section*{References} 

\bibliographystyle{iopart-num}
\bibliography{refs}

\providecommand{\newblock}{}
\begin{thebibliography}{10}
\expandafter\ifx\csname url\endcsname\relax
  \def\url#1{{\tt #1}}\fi
\expandafter\ifx\csname urlprefix\endcsname\relax\def\urlprefix{URL }\fi
\providecommand{\eprint}[2][]{\url{#2}}

\bibitem{Harry:2010}
Harry G~M and the LIGO Scientific~Collaboration 2010 {\em Classical and Quantum
  Gravity\/} {\bf 27} 084006

\bibitem{Iyer:2011wb}
Iyer B, Souradeep T, Unnikrishnan C, Dhurandhar S, Raja S, Kumar A and Sengupta
  A~S 2011 {LIGO-India} Tech. rep.
  \urlprefix\url{https://dcc.ligo.org/cgi-bin/DocDB/ShowDocument?docid=75988}

\bibitem{adV}
Advanced {V}irgo \url{http://wwwcascina.virgo.infn.it/advirgo/}

\bibitem{kagra}
{KAGRA} \url{http://gwcenter.icrr.u-tokyo.ac.jp/en/}

\bibitem{lofar}
{LOFAR} \url{http://www.lofar.org/}

\bibitem{Rau:2009yx}
Rau A, Kulkarni S~R, Law N~M, Bloom J~S, Ciardi D, Djorgovski G~S, Fox D~B,
  Gal-Yam A, Grillmair C~C, Kasliwal M~M, Nugent P~E, Ofek E~O, Quimby R~M,
  Reach W~T, Shara M, Bildsten L, Cenko S~B, Drake A~J, Filippenko A~V, Helfand
  D~J, Helou G, Howell D~A, Poznanski D and Sullivan M 2009 {\em Publications
  of the Astronomical Society of the Pacific\/} {\bf 121} pp. 1334--1351 ISSN
  00046280

\bibitem{panstarrs}
Pan-{STARRS} \url{http://pan-starrs.ifa.hawaii.edu/public/}

\bibitem{skymapper}
Sky {M}apper \url{http://rsaa.anu.edu.au/skymapper/}

\bibitem{nakar07}
Nakar E 2007 {\em Phys. Rept.\/} {\bf 442} 166--236

\bibitem{Ott}
Ott C~D 2009 {\em Classical and Quantum Gravity\/} {\bf 26} 204015

\bibitem{Schutz:1986gp}
Schutz B~F 1986 {\em Nature\/} {\bf 323} 310--311

\bibitem{Schutz:2011fn}
Schutz B~F 2011 {\em Classical and Quantum Gravity\/} {\bf 28} 125023

\bibitem{GMRT}
Giant metrewave radio telescope \url{http://gmrt.ncra.tifr.res.in/}

\bibitem{Iyer:2011}
Iyer B {\em Private Communication\/}

\bibitem{LIGOSouth}
Klimenko S, Saulson P, Sathyaprakash B, Fritschel P, Raab F, Weiss R and Finn
  L~S 2010
  \urlprefix\url{https://dcc.ligo.org/cgi-bin/DocDB/ShowDocument?docid=11604}

\bibitem{Veitch:2012df}
Veitch J, Mandel I, Aylott B, Farr B, Raymond V {\em et~al.\/} 2012
  (\textit{Preprint} \eprint{1201.1195})

\bibitem{Fairhurst:2009tc}
Fairhurst S 2009 {\em New J. Phys.\/} {\bf 11} 123006

\bibitem{Vitale:2010fg}
Vitale S and Zanolin M 2010 {\em Physical Review D\/} {\bf 82} 124065

\bibitem{vanderSluys:2007st}
van~der Sluys M, Roever C, Stroeer A, Christensen N, Kalogera V {\em et~al.\/}
  2007  (\textit{Preprint} \eprint{0710.1897})

\bibitem{Virgo:2011aa}
{Abadie} J, {Abbott} B~P, {Abbott} R, {Abbott} T~D, {Abernathy} M, {Accadia} T,
  {Acernese} F, {Adams} C, {Adhikari} R, {Affeldt} C and et~al 2012 {\em Astron
  Astrophys\/} {\bf 541} A155

\bibitem{aligo}
 2009 Advanced {LIGO} anticipated sensitivity curves
  \urlprefix\url{https://dcc.ligo.org/cgi-bin/DocDB/ShowDocument?docid=2974}

\bibitem{Fairhurst:2010is}
Fairhurst S 2011 {\em Class.Quant.Grav.\/} {\bf 28} 105021

\bibitem{2012PhRvD..85h2002A}
Abadie J {\em et~al.\/} 2012 {\em Physical Review D\/} {\bf 85} 82002

\bibitem{Abadie:2010cfa}
Abadie J {\em et~al.\/} 2010 {\em Class. Quant. Grav.\/} {\bf 27} 173001

\bibitem{Reitze:Bf7sDr2U}
Sathyaprakash B, Fairhurst S, Schutz B, Veitch J, Klimenko S, Reitze D and
  Whitcomb S {Scientific benefits of moving one of LIGO Hanford detectors to
  India} Tech. rep.
  \urlprefix\url{www.ligo.caltech.edu/NSF/related/10.2011/ligo-india-110923.pdf}

\end{thebibliography}

\end{document}